\documentclass[DIV=12,headsepline=true]{scrartcl}
\usepackage{amsmath,amssymb}
\usepackage[
colorlinks=true,
linkcolor=blue,
citecolor=blue,
urlcolor=blue,
linktocpage=true
]{hyperref}
\usepackage[capitalise]{cleveref}
\usepackage{slashed}
\usepackage{bm}
\usepackage{braket}
\usepackage{ascmac}
\usepackage{mathrsfs}
\usepackage{cite}
\usepackage{graphicx}
\usepackage{mathtools}
\usepackage{color}

\usepackage[inline,final]{showlabels}
\showlabels{cite}
\showlabels{ref}
\showlabels{cref}

\newcommand{\hodge}{{\star}}
\newcommand{\tr}{\operatorname{tr}}

\crefname{section}{Sec.}{Secs.}

\begin{document}

\begin{titlepage}
\setkomafont{pagenumber}{\normalfont\color{white}}
\setkomafont{pageheadfoot}{\normalfont\normalcolor}
\thispagestyle{myheadings}
\markright{\rightline{KEK-TH-2313}}
\centering

\vspace*{100pt}
{\usekomafont{disposition}\huge Missing final state puzzle in the
monopole-fermion scattering}

\renewcommand{\thefootnote}{\fnsymbol{footnote}}

\vspace{30pt}
{\Large Ryuichiro Kitano$^{1,2}$ and Ryutaro Matsudo$^{1}$}

\renewcommand{\thefootnote}{\arabic{footnote}}
\setcounter{footnote}{0}

\vspace{10pt}

\textit{\large ${}^{1}$KEK Theory Center, Tsukuba 305-0801, Japan}

\textit{\large ${}^{2}$Graduate University for Advanced Studies (SOKENDAI), Tsukuba 305-0801, Japan}

\vspace{10pt}
{\large \textit{E-mail}: \href{mailto:ryuichiro.kitano@kek.jp}{\texttt{ryuichiro.kitano@kek.jp}}, \href{mailto:ryutaro.matsudo@kek.jp}{\texttt{ryutaro.matsudo@kek.jp}}}
\vspace{30pt}

{\usekomafont{disposition} \large Abstract}
\vspace{0pt}

\begin{abstract}
It has been known that when a charged fermion scatters off a monopole, the fermion in the $s$-wave component must flip its chirality, i.e., fermion number violation must happen. This fact has led to a puzzle; if there are two or more flavors of massless fermions, any superposition of the fermion states cannot be the final state of the $s$-wave scattering as it is forbidden by conservation of the electric and flavor charges. The unitary evolution of the state vector, on the other hand, requires some interpretation of the final states.  We solve the puzzle by finding new particle excitations in the monopole background, where multi-fermion operators exhibit condensation.  The particles are described as excitations of closed-string configurations of the operators.


\end{abstract}

\end{titlepage}

\setcounter{page}{1}

\tableofcontents
\noindent\hrulefill

\section{Introduction}
It has been known that an 't~Hooft-Polyakov
monopole~\cite{Hooft1974a,Polyakov1974} can catalyze proton decays in
grand unified theories (GUTs) at the rate of the typical strong
interaction~\cite{Rubakov1982,Callan1982,Callan1983}. The origin of
this effect is that, when a single charged fermion collides with a
monopole, the helicity of the lowest partial wave of the fermion has
to flip~\cite{Kazama1977}. 
This forces the $s$-wave component of the incoming proton to interact
with the monopole core where the GUT gauge interaction is not
suppressed.
The higher partial waves are kept away from the monopole core
according to the analysis of the Dirac equation in the monopole
background~\cite{Jackiw1976}. The process of the fermion-monopole
scattering was analysed in the $s$-wave approximation where the fields
are projected into the spherically symmetric component~\cite{Rubakov1982,Callan1982,Callan1983}. In this approximation, the
theory reduces to a Schwinger model, and thus when the fermion is
massless, the $s$-wave theory can be solved exactly. In the massive
case, the bosonization of the theory is useful, where the fermions are
described as kink solitons. A numerical study of the scattering problem has been performed by solving the equation of motion in the bosonized theory~\cite{Dawson1983}.

The soliton picture of the scattering can also be adopted when the
effect of confinement is included by regarding a proton as a skyrmion
\cite{Callan1984a}. In this framework, the skyrmion decays into a
positron if the boundary condition at the core of the monopole
violates the baryon number conservation.

Recently, scattering amplitudes including magnetically charged particles are reconsidered using the on-shell method
\cite{Csaki2020}.
A multi-particle state including a magnetically charged particle and an electrically charged one simultaneously cannot be written as a direct product of one-particle state because the state has an additional quantum number called the pairwise helicity, which is the ``cross product'' of the electric and magnetic charges.
The helicity flip of the lowest partial wave can also be derived using this technique.

Although the ability of monopole to catalyze proton decay was proposed
a long time ago, there remains a puzzle; when the incoming particle is
a massless fermion in a theory with two or more flavors of massless Dirac
fermions, there seems to be no final state consistent with all the
conservation laws~\cite{Sen1983,Callan1984,Preskill1984,Rubakov1988}.
In the $s$-wave calculation, the final state consists of objects with fractional fermion numbers, which makes it difficult to interpret what it actually is. 

In this paper, we consider an $SU(2)$ gauge theory with massless
fermions and 't~Hooft-Polyakov monopoles. We resolve the puzzle by
finding new fermionic states in the system.
In the monopole background, one can identify the final states as
soliton excitations that are ``new'' fermions with
opposite helicity to the original fermions.
In this picture, the fermion condensation around monopole plays a
crucial role. Both new and original fermions can be described as the
solitons of the multi-fermion operators that exhibit condensation without adopting the $s$-wave
approximation, i.e., in four space-time dimensions.
They are walls bounded by strings around which the phases of the
multi-fermion operators wind. We call them ``pancakes.'' The helicity flip
in the scattering is understood by following the
classical time evolution of the pancakes.

For $N_f\geq 2$, a pancake with the opposite helicity cannot be
interpreted as one of the original fermions because its flavor charge
does not match. This means that when there is a monopole, there are
one-particle states that are not the original fermions.
When we add masses to the fermions, the soliton that comes from the
monopole after the scattering can only be considered as an
intermediate state since it is no longer stable. The soliton further
deforms into one of the original fermions. 

%

The pancake solitons we consider in this paper are similar to the
$\eta'$ strings, which are proposed to describe spin-$N_c/2$ baryons
in large $N_c$ QCD~\cite{Komargodski2018a}. In the case of $\eta'$,
which represents the phase of the quark bilinear, the pancake is
shown to support the Chern-Simons theory which has a dynamical mode on
the edge. The quantization of the edge mode provides baryon states
with appropriate quantum numbers~\cite{Komargodski2018a, Kitano2021}.
Our discussion is parallel to this. The phases of the multi-fermion
operators around the monopole can have pancake-like topologically
non-trivial configurations. 
The pancakes interact with the photon via the Chern-Simons coupling, which implies the existence of the chiral edge modes on the boundary.
The excitations of the edge modes describe the fermions in the theory.
%
A closely related discussion has been given in QED coupled with an
axion. The axion string in the theory has similar properties as
discussed in Ref.~\cite{Kogan1993}.

In the following section, we review the puzzle in the monopole-fermion
system in the massless limit. The new states to account for the final
states of the scattering are identified in Section~\ref{sec:solitons},
and the scattering processes are considered in the soliton picture in
Section~\ref{sec:scattering}. We summarize the discussion in
Section~\ref{sec:summry}.


\section{Semiton puzzle}
\begin{figure}
\centering
\includegraphics[width=0.6\hsize]{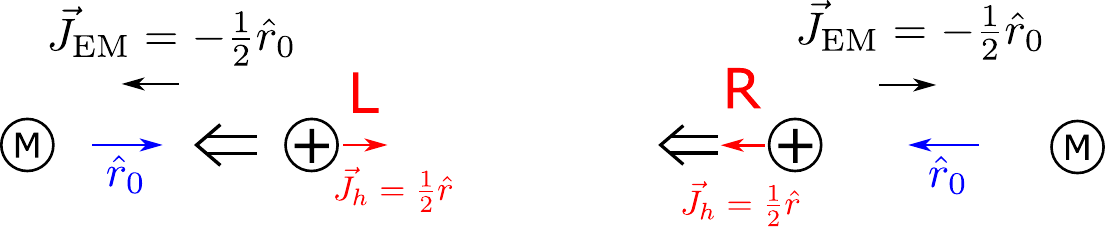}
\caption{The angular momentum when a left-handed fermion with positive unit charge goes into a monopole with zero impact parameter.}
\label{fig:s_wave}
\end{figure}

We consider the $SU(2)$ gauge theory coupled with an adjoint scalar
$\Phi$ and $2N_f$ Weyl fermions $\chi_j$ in the fundamental
representation, whose Lagrangian is
\begin{align}
  &\mathcal L = -\frac1{4g^2} 2\tr(\mathsf f\hodge \mathsf f) + i\sum_{j=1}^{2N_f}\overline\chi_j\,\overline\sigma^\mu D_\mu\chi_j + \frac12 2\tr(D\Phi\hodge D\Phi) - \frac{\lambda}4(2\tr\Phi^2-v^2)^2,\notag\\
  &\mathsf f = d\mathsf a -i \mathsf a^2 = f^a \sigma^a/2,\quad \Phi = \Phi^a \sigma^a/2,\quad D\Phi = d\Phi -i[\mathsf a,\Phi],\quad D\chi_j = (d-i\mathsf a)\chi_j,
\end{align}
where $\mathsf a$ and $\mathsf f$ are the $SU(2)$ gauge field and its field strength respectively.
We set the potential for the adjoint scalar so that it has a non-zero
expectation value. The gauge group $SU(2)$ is spontaneously broken
down to $U(1)$. Note that this theory has $SU(2N_f)$ flavor symmetry
rather than $SU(N_f)\times SU(N_f)$ because of the pseudo-reality of
the $SU(2)$ gauge group.
The symmetry plays an important role in the following discussion. We
denote the components of the left-handed Weyl fermions, $\chi_j$, as
$a_j$ and $b_j$, whose charges under the unbroken $U(1)$ group are
$1$ and $-1$, respectively.
The fermions $a_j$ and $b_j$ have $2N_f$ components which form the
fundamental representation under the $SU(2N_f)$ flavor group. At low
energy, the theory reduces to the $U(1)$ gauge theory as
\begin{align}
  &\mathcal L_{\mathrm{eff}} = -\frac1{4g^2} f\hodge f + i\overline{a_j}\,\overline\sigma^\mu D_\mu a_j + i\overline{b_j}\,\overline\sigma^\mu D_\mu b_j,\notag\\
  &D a_j = (d-ia)a_j ,\quad Db_j = (d+ia)b_j,
  \label{eff}
\end{align}
where $a$ without index and $f$ are the $U(1)$ gauge field and its field strength respectively.
Let us consider what happens when a particle with a unit charge collides with an 't~Hooft-Polyakov monopole. 
Even if the particles are at rest, the $U(1)$ gauge field (that is the electromagnetic field) carries an angular momentum
\begin{align}
  \vec J_{\mathrm{EM}} = \frac1{4\pi}\int d^3x\,{\vec r}\times(\vec E\times\vec B) = -\frac12\hat{r}_0,
  \label{a_mom}
\end{align}
where $\hat{r}_0$ is the unit vector pointing from the magnetic monopole to the charge.
Therefore, if the particle has a helicity $-1/2$ (left-handed) and the impact parameter is zero, the total angular momentum is zero.
After the collision, the angular momentum from the $U(1)$ gauge field has the opposite direction to the particle momentum.
See \cref{fig:s_wave}.
This means that, in order to maintain the total angular momentum to be zero, the helicity of the particle must flip
\cite{Kazama1977,Rubakov1982,Callan1982,Callan1983}.
This causes a problem when $N_f\geq 2$.
The helicity, the $U(1)$ charge and the representation of $SU(2N_f)$ of the fermions are
\begin{align}
  \bm a:\ (L,+1,\Box),\quad \bm b:\ (L,-1,\Box),\quad \overline{\bm a}:\ (R,-1,\overline\Box),\quad \overline{\bm b}:\ (R,+1,\overline\Box).
\end{align}
However, if the incoming particle is $\bm a$, there are no corresponding outgoing particles because the quantum numbers of the outgoing particle must be $(R,+1,\Box)$ for the conservation of the total angular momentum, the $U(1)$ charge, and the $SU(2N_f)$ charge.


When $N_f=2$ and the incoming state is the $s$-wave of $a_1$, the conservation of the charge and the flavor quantum number implies that the final state is something like
\begin{align}
  \frac{b_1}2 + \frac{\overline b_2}2 + \frac{\overline b_3}2 + \frac{\overline b_4}2,
  \label{fs}
\end{align}
where $b_j/2$ ($\overline b_j/2$) denotes the state whose quantum numbers are halves of those of $b_j$ ($\overline b_j$).
The state $b_j/2$ is sometimes called as a ``semiton''~\cite{Preskill1984}.

In Ref.~\cite{Callan1984,Rubakov1988}, the semiton state is interpreted as follows.
If we regard the theory as an approximation of the $SU(5)$ GUT with a monopole, the correspondence of the particles is
\begin{align}
  a_1 = e^+_L,\quad a_2=\overline d^3_L,\quad a_3 = u^1_L,\quad a_4 = u^2_L, \notag\\
  b_1 = d^3_L,\quad b_2= e^-_L,\quad b_3 =\overline u^2_L,\quad b_4 =\overline u^1_L.
\end{align}
The final state (\ref{fs}) is interpreted as the superposition of $e^+_R$ and $u^1_Ru^2_Rd^3_L$ with equal weights.
However, this is problematic because the $SU(2N_f)$ charge is not conserved.
Note that $e_R^+=\overline b_2$ is in the antifundamental representation of $SU(2N_f)$.

We argue that the state (\ref{fs}) should be interpreted as a ``new
particle'', whose quantum number is $(R,+1,\Box)$.
The new particle is not a single excitation from the vacuum. It is
rather a part of a two-particle state: a monopole and a new particle,
but the new particle can be arbitrarily far away from the monopole. We
will be able to find such a state as a soliton of the multi-fermion
operators around the monopole.


The puzzle disappears when we add fermion masses since the flavor
symmetry is reduced. An appropriate final state can always be found as
(a combination of) the original fermions.
Therefore, one can understand the Callan-Rubakov effects as a simple
scattering process in the massive case. The final states are
particles. The discussion of the massless limit should recover this
result when masses are added. We will discuss how the massless limit
and the massive case are connected in Section~\ref{sec:scattering}.

It was confirmed that the final state has the fractional fermion numbers by applying the s-wave projection \cite{Maldacena1997,Dawson1983,Callan1984,Rubakov1988}.
The $s$-wave projection is performed by substituting the spherically symmetric fields, whose total angular momentum is zero.
At low energy, the theory reduces to the $2N_f$-flavor QED in two dimensions with the space-dependent coupling.
In the bosonized theory \cite{Mandelstam1975,Coleman1975}, the scalar fields correspond to the phases of the fermion bilinears, and the fermions are described as a kink soliton made of the phases.
By solving the equation of motion, we see that the final state corresponds to a kink with fractional fermion numbers \cite{Dawson1983,Rubakov1988}.
This theory for $N_f=2$ in $g\to0$ limit was analysed in Ref.~\cite{Maldacena1997}, and it was found that the theory contains new fermions with the opposite $U(1)_A$ charge as the original fermion in the action as a collective mode of the fermion fields.

The analysis in the s-wave projection is robust for the following reason.
We can safely take $g\to0$ limit to consider the puzzle because the helicity flip and the charge conservation is maintained.
In the limit, the fermions freely propagate when it does not reach the core of the monopole.
From the analysis of the Dirac equation, we see that the higher partial waves cannot reach the core due to the centrifugal barrier term
\cite{Jackiw1976,Callan1982}, and thus no mixing between the s-wave and the higher partial wave can occur.
Therefore, the analysis in the s-wave projection is exact in the $g\to0$ limit.
The boundary condition that is consistent with the flavor and electric charge conservation is uniquely determined in the s-wave theory, which gives the final state of the scattering with the fractional fermion numbers.
The only remaining problem is what the interpretation of the final state is.
We give an answer to this problem by giving a four dimensional picture that reproduces the two dimensional result when we restrict ourselves to the spherically symmetric system.

\section{Fermions as solitons}
\label{sec:solitons}

Around the monopoles, there are condensations of multi-fermion
operators which break the anomalous $U(1)$ symmetry while the
$SU(2N_f)$ global symmetry is left unbroken~\cite{Rubakov1982,
Craigie1984}.
We will describe in this section how the configurations of the multi-fermion operator can be identified with the fermions. We will find that
there are configurations which have the same quantum numbers as the
original fermions, $a_j$ and $b_j$, as well as new fermion states
which are to be identified as the final state of the scattering
processes.


We consider the configuration where there is a static
't~Hooft-Polyakov monopole, which we approximate as a Dirac monopole
for the $U(1)$ gauge field. 
Around the monopole, operators made out of the fermion fields have
nonzero expectation values. The possible condensates are
\cite{Kazama1983,Rubakov1988}
\begin{align}
  \Braket{(a_{i_1}b_{i_2})\cdots (a_{i_{2N_f-1}}b_{i_{2N_f}})}= \frac1{r^{3N_f}}c \varepsilon_{i_1\ldots i_{2N_f}},
  \label{condensate2}
\end{align}
where $r$ is the distance from the monopole, and $c$ is a constant.
We can consider string configurations around which the phases of
these operators wind. 
As we will explain below, the strings can have
nonzero electric charges, spin half and $SU(2N_f)$ charges, which enable
us to regard the strings as fermions.
This string configurations are similar to the axion strings, which can also have nonzero charge and spin half \cite{Callan1985}.
In general, such a string configuration exists when the field that is transformed under $U(1)_A$ has a nonzero expectation value.

It is possible that the condensate (\ref{condensate2}) affects the asymptotic states even though the value goes to zero at spacial infinity.
To see what happens, let us first consider the s-wave
theory. In the s-wave theory, there is a collective mode of the
fermionic field corresponding to the semiton state, which is described
as a kink in the bosonized picture.
In our view, this kink solution should be lifted up to four dimensions, i.e., to a
large wall surrounding the monopole. This is one of the examples of what
happens in the $r \to \infty$ limit, meaning that the kinetic energy of
the scalar (i.e. the phase of the operator) is preserved by expanding
the size of the wall even if the value of the condensate is
reduced. This large wall should be regarded as a spherically symmetric
wave function of the fermion, given its correspondence to a kink in the
s-wave theory.
Next, let us consider the case of a pancake, a wall bounded by a string. Since we want to consider a
pancake as a particle, it should be possible to make it smaller at any
point in space. In the case of a pancake, there is another way to
maintain its kinetic energy than to increase its size. That is to become
thinner as it moves away from the monopole. 
Although we did not try to find an explicit pancake solution as we do
not have a complete effective theory of condensates, the interpretation
of the pancake as a particle is at least consistent.

\subsection{Monopole bags and the Witten effect}
To see the effect of the condensates
(\ref{condensate2}), we try to determine the effective Lagrangian that
is obtained by ``integrating in'' their phases. 
A phase shift of the
operators is induced by a combination of phase shifts of the
fermions, which gives a shift of the Lagrangian due to the chiral
anomaly. 
Below, we consider an effective theory where the phases of the
fermions are the dynamical degrees of freedom.
%
%
Let $\alpha_j$ and $\beta_j$ be the phases of $a_j$ and $b_j$,
respectively. Note that there are configurations of $\alpha_j$ and
$\beta_j$ that represent an identical configuration of the phases of the multi-fermion operators, which means there is a redundancy of description. This is
because, under $SU(2N_f)$ transformations and $U(1)$ gauge
transformations, $\alpha_j$ and $\beta_j$ change but the multi-fermion operators
do not.

We can consider the domain wall configuration of $\alpha_j$ and $\beta_j$ because they are $2\pi$ periodic.
When the wall wraps the monopole, the $U(1)$ charge and the $SU(2N_f)$ charge are induced due to the Witten effect~\cite{Witten1979c}.
We call such an object as a monopole bag~\cite{Kogan1993}.
To determine the charges of a monopole bag, we consider the background gauge field for the $U(1)$ subgroups of $SU(2N_f)$.
The covariant derivative of $a_j$ is
\begin{align}
  (d-ia -iA_l[H_l]_{jj})a_j,
\end{align}
where $A_l$ is the gauge
field corresponding to a Cartan generator $H_l$ in
$\mathfrak{su}(2N_f)$, and $[M]_{ij}$ denotes the $ij$-component of
the matrix $M$. The phase rotations give the spacetime dependent
``$\theta$ terms,''
\begin{align}
  \frac1{8\pi^2}\sum_j \left( \alpha_j(f + F_l[H_l]_{jj})^2 + \beta_j(-f + F_l[H_l]_{jj})^2\right).
  \label{topo_term}
\end{align}
The boundary condition at the core of the monopole is\footnote{
  The right-hand side of the conditions (\ref{boundary_cond}) is not integer multiples of $2\pi$ but zero.
    When we replace the monopole by a dyon, the right-hand side changes to $2\pi$ times a corresponding integer so that the charge of the dyon is canceled by the charge coming from the Witten effect.
}
\begin{align}
  &\theta_A := \sum_j (\alpha_j + \beta_j) = 0,\notag\\
  &\theta_{jk}:= \alpha_j - \beta_j - \alpha_k + \beta_k = 0, \quad \forall j,k.
  \label{boundary_cond}
\end{align}
This condition is determined so that there is no divergence of the action other than that of the kinetic term $-f\hodge f/4$ corresponding to the mass of the monopole.
The divergence should vanish for any value of the background gauge fields.
This boundary condition ensures the conservation of the electric and flavor charges.
Since the phases of the multi-fermion operators in Eq.~(\ref{condensate2}) can be expressed as linear combinations of $\theta_A$ and $\theta_{jk}$, the condition (\ref{boundary_cond}) means that the phases of the condensates has to be fixed.
From this expression, we can read off the $U(1)$ current $J$ and the currents $J_l$ for the maximal torus of $SU(2N_f)$.
If $A_l =0$, they are
\begin{align}
  &\hodge J = \frac1{4\pi^2} d\theta_A f \notag\\
  &\hodge J_l = \sum_j [H_l]_{jj}\frac1{4\pi^2}d(\alpha_j-\beta_j).
  \label{currents}
\end{align}
Therefore, the monopole bag corresponding to the positive $2 \pi$
shift of $\alpha_j$, i.e., $2\pi$ shifts of $\theta_A$ and $\theta_{jk}$, has the $U(1)$ charge $Q=1$ and the same $SU(2N_f)$ charges as $a_j$.
Also, the monopole bag correspond to the $2\pi$ shift of $\beta_j$ has the same charges as $b_j$.
%
By inserting a thin wall where $\alpha_j$ changes from $0$ to $2\pi$ to the topological term (\ref{topo_term}), we obtain the Chern-Simons coupling with the level unity on the wall
\begin{align}
  \frac1{4\pi}\int_{M^3} (a+A_l[H_l]_{jj})(f+F_l[H_l]_{jj}),
  \label{CS}
\end{align}
where $M^3$ is the world volume of the wall.

\subsection{Pancake soliton}
Next, let us consider the string configuration around which $\alpha_j$
winds. Such an object exists because at the core of the string, the
values of the multi-fermion fields can be zero so that the configuration is smooth everywhere. There is a domain wall bounded by the string,
which means that the object is considered as the monopole bag with a
hole. As we mentioned above, there is the Chern-Simons coupling 
(\ref{CS}) on the wall. Thus there should be a chiral edge mode on the
string because otherwise the gauge symmetry is broken.
Let us consider the case where the domain wall is a two-dimensional
disc $D^2$. The theory on the wall bounded by the string is~\cite{Elitzur1989}
\begin{align}
  &\frac1{4\pi}\int_{\mathbb R\times D^2} (a+A_l[H_l]_{jj})(f+F_l[H_l]_{jj}) \notag\\
  &+ \frac1{4\pi}\int_{\mathbb R\times \partial D^2}(D_x\phi (D_t\phi+v D_x\phi)\, dxdt - \phi(f+F_l[H_l]_{jj})),
\end{align}
where $x$ is a coordinate of $\partial D^2$ that is periodic and
$\phi$ is a $2\pi$-periodic scalar on $\partial D^2$, that transforms
as $\phi\rightarrow \phi + \lambda + \lambda_l[H_l]_{jj}$ under the
gauge transformations $a\rightarrow a+d\lambda$, $A_l\rightarrow A_l +
d\lambda_l$. The covariant derivative is defined as $D\phi :=
d\phi-a-A_l[H_l]_{jj}$, and $v$ is a constant. The $U(1)$ charge $Q$
and $SU(2N_f)$ charges $Q_l$ can be written as\footnote{In two
dimension, the definition of the current $j$ depends on the
``position'' of $a$ in the integral when taking the variation,
\begin{align}
  \int \hodge j a = -\int a\hodge j =\int a \hodge j', \Rightarrow j = -j'.
\end{align}
In this paper, we define $j$ by putting $a$ to the right.
We should note this if one combines the charges from the edge and the bulk.
}
\begin{align}
  &Q =\frac1{2\pi} \int_{\partial D^2}( d\phi - a - A_l[H_l]_{jj}) + \frac1{2\pi} \int_{D^2} (f+F_l[H_l]_{jj}), \notag\\
  &Q_l =\frac1{2\pi} \int_{\partial D^2}( d\phi - a - A_k[H_k]_{jj})[H_l]_{jj} + \frac1{2\pi} \int_{D^2} (f+ F_k [H_k]_{jj})[H_l]_{jj}.
  \label{pancake_charge}
\end{align}
If we take the gauge where the Dirac string from the monopole does not
penetrate the wall, there is no contribution from the gauge fields $a$
or $A_l$ due to the cancellation between the bulk and edge
contributions. As a result, the charge of the object is identified
with the winding number of $\phi$. Note that, as the monopole becomes
closer to the wall, the electric charge density moves from the edge to
the bulk because the contribution from $a$ partially cancels that from
$\phi$ on the edge, and $f$ contributes to the charge in the bulk.
The local operator that creates the state with charge $\pm 1$ is
\begin{align}
  e^{\mp i\phi},
  \label{state}
\end{align}
which is confirmed by its commutator with the charge operator $\int
\partial_x\phi\, dx/(2\pi)$ for $A=a=0$. The spin of the state can be
calculated as the eigenvalue of $LP_x/(2\pi)$, where $L$ is the
circumference of $D^2$ and $P_x:=\int (\partial_x\phi)^2\,dx/(4\pi)$
is the generator of the translation along $x$. The component
perpendicular to $D^2$ of the spin of the state (\ref{state}) is found to be
$1/2$. The direction of the spin depends only on the
orientation of the domain wall, irrespective of the sign of the
charge.
\begin{figure}[t]
\centering
\includegraphics[width=0.7\hsize]{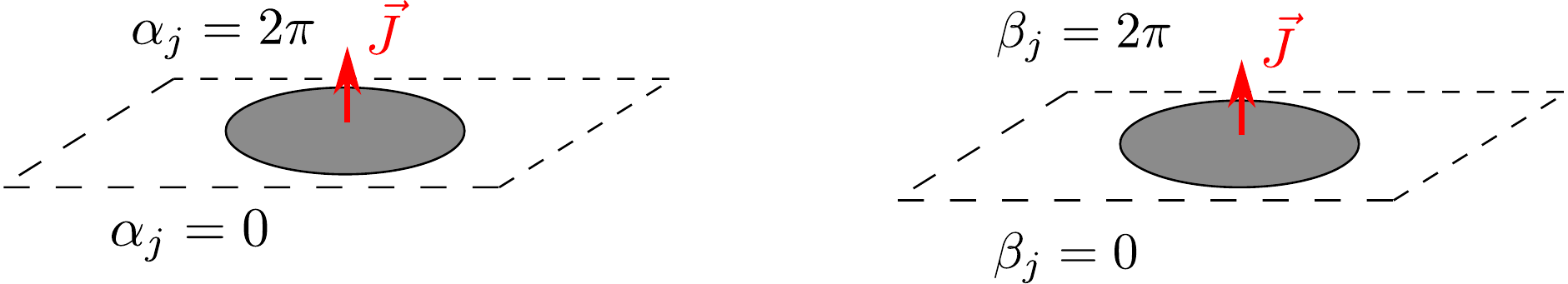}
\caption{The direction of the spin of the pancake. The red arrow denotes the direction of the spin.}
\label{fig:spin}
\end{figure}
%
%
See \cref{fig:spin}.

The theory on the wall bounded by the string for $\beta_j$ is similar to that for $\alpha_j$ with the exception of the sign in front of $A_l$:
\begin{align}
  &\frac1{4\pi}\int_{\mathbb R\times D^2} (a-A_l[H_l]_{jj})(f-F_l[H_l]_{jj}) \notag\\
  &+ \frac1{4\pi}\int_{\mathbb R\times\partial D^2}(D_x\tilde \phi (D_t\tilde \phi+v D_x\tilde \phi)\, dxdt - \phi(f-F_l[H_l]_{jj})).
\end{align}
The covariant derivative is defined as $D\tilde\phi = (d-ia+iA_l[H_l]_{jj})\tilde\phi$.

\begin{figure}[t]
\centering
\includegraphics[width=0.70\hsize]{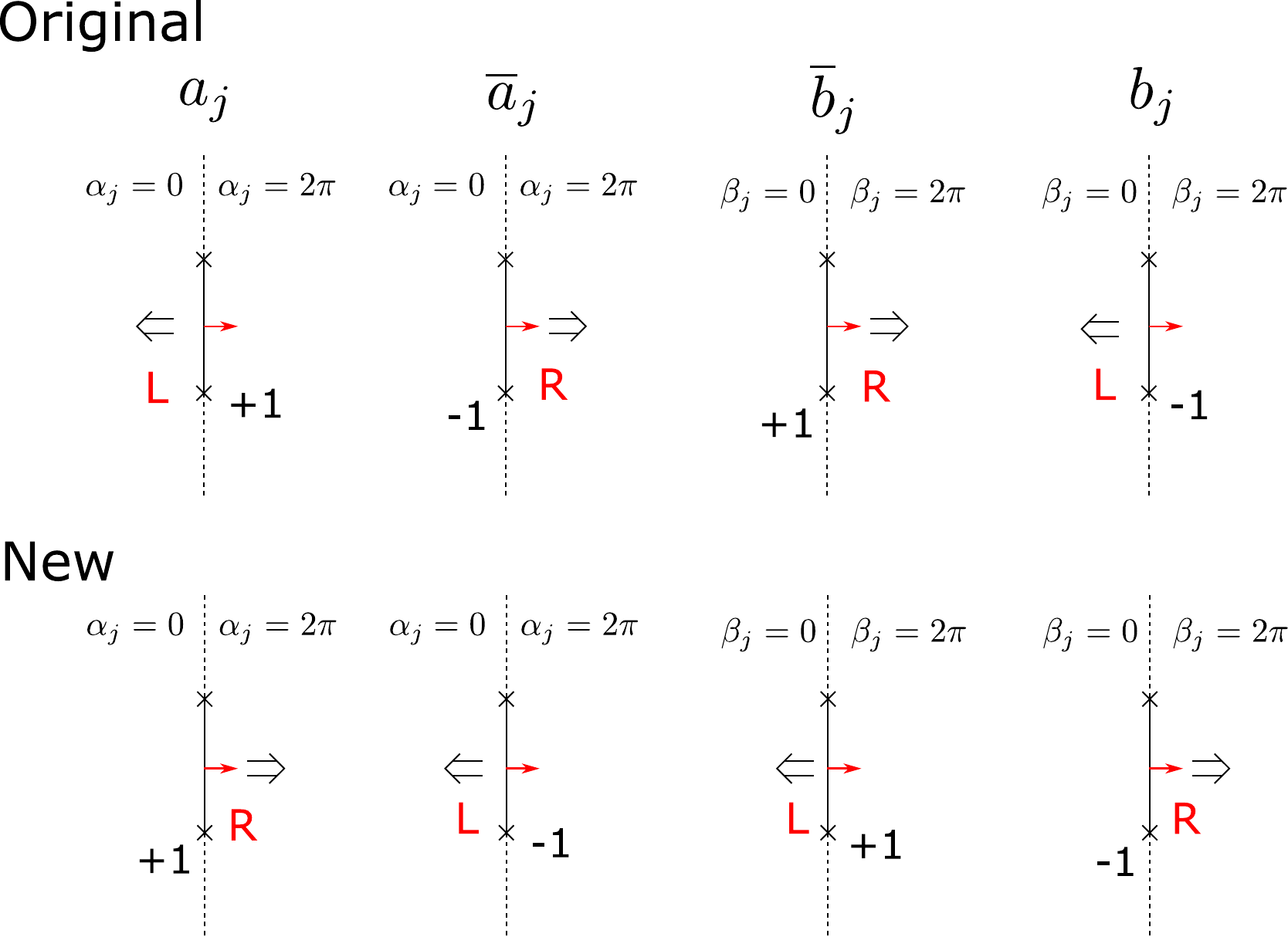}
\caption{Pancakes corresponding to the original fermions (upper panel) and the new fermions (bottom panel). The double arrows denote the direction of the motion. The red arrows denote the direction of the spin.}
\label{fig:pancakes}
\end{figure}

The pancakes have unit charge, unit flavor charge and
spin $1/2$. Therefore we can regard them as fermions. However, the
original fermions $a_j$ and $b_j$ correspond to the pancakes with an
appropriate direction of motion, and those moving in the opposite
direction should be regarded as new particle states. See
\cref{fig:pancakes}.
We will see in the next subsection that one of them is the final state
of the monopole scattering when the initial state is one of $a_j$ and
$\bar a_j$.

The boundary of the pancake cannot just shrink because of the conserved quantum numbers.
However, when the wall surrounds the monopole it gets possible, and thus the pancake can continuously deform to a monopole bag.
When the pancake wraps the monopole, the charge at the edge moves toward the bulk as we can see in Eq.~(\ref{pancake_charge}), then finally there are no charge at the edge and it can shrink.

Since we find a stable configuration of the phases, it just
exists. Although it sounds surprising to find such a new state, the
existence has already been admitted in the analyses of the $s$-wave
reduced two-dimensional theory. In the bosonized picture, there are
solitons which can be the final state of the scattering. We argue that
the soliton state in the two-dimensional analyses is simply promoted
to be a particle state in four dimensions as it should be.

Because the (original and new) fermions are massless, the pancake has to be a massless soliton, which has different features than a usual massive soliton.
Since a massless soliton cannot be static by definition, we cannot define the tension, which is defined using the energy density of the static solution.
Correspondingly, the potential term should vanish.
For example, let us consider the
bosonized version of the free fermion theory in two dimension. In the
theory, a fermion is described as a sine-Gordon kink, where the fermion
mass term can be regarded as a potential for the scalar field. In the
massless limit of fermions, the potential disappears while the massless
fermions are still described as kinks. The kink solution can no longer
be static, reflecting the fact that fermions move at the speed of
light. Then we see that a potential is needed
for the existence of the static soliton solution, which cannot be
massless, while the potential should vanish for the existence of a
massless soliton, which cannot be static.
In our case, the pancake
describes the massless fermion, then the potential should vanish. 
[The
interaction with photon gives the potential corresponding to the Coulomb
interaction, but it can be neglected when we consider an isolated
fermion.]

\subsection{Soliton picture of the scattering}
\label{sec:scattering}

By identifying the fermions $a_j$ and $b_j$ as solitons, one can
discuss the scattering with the monopole as classical time evolutions
of the solitons. We will see how the initial state deforms into the
new fermion states. We also discuss what happens when fermion masses
are added.

In this section, we set $N_f=2$ and consider the situation where a
pancake hits the monopole. For concreteness, we take the initial state
as the charge-one excited pancake made of $\alpha_1$, i.e., $a_1$. The
corresponding situation for the axion strings is considered in
Ref.~\cite{Kogan1993}.

\begin{figure}[t]
\centering
\includegraphics[width=0.70\hsize]{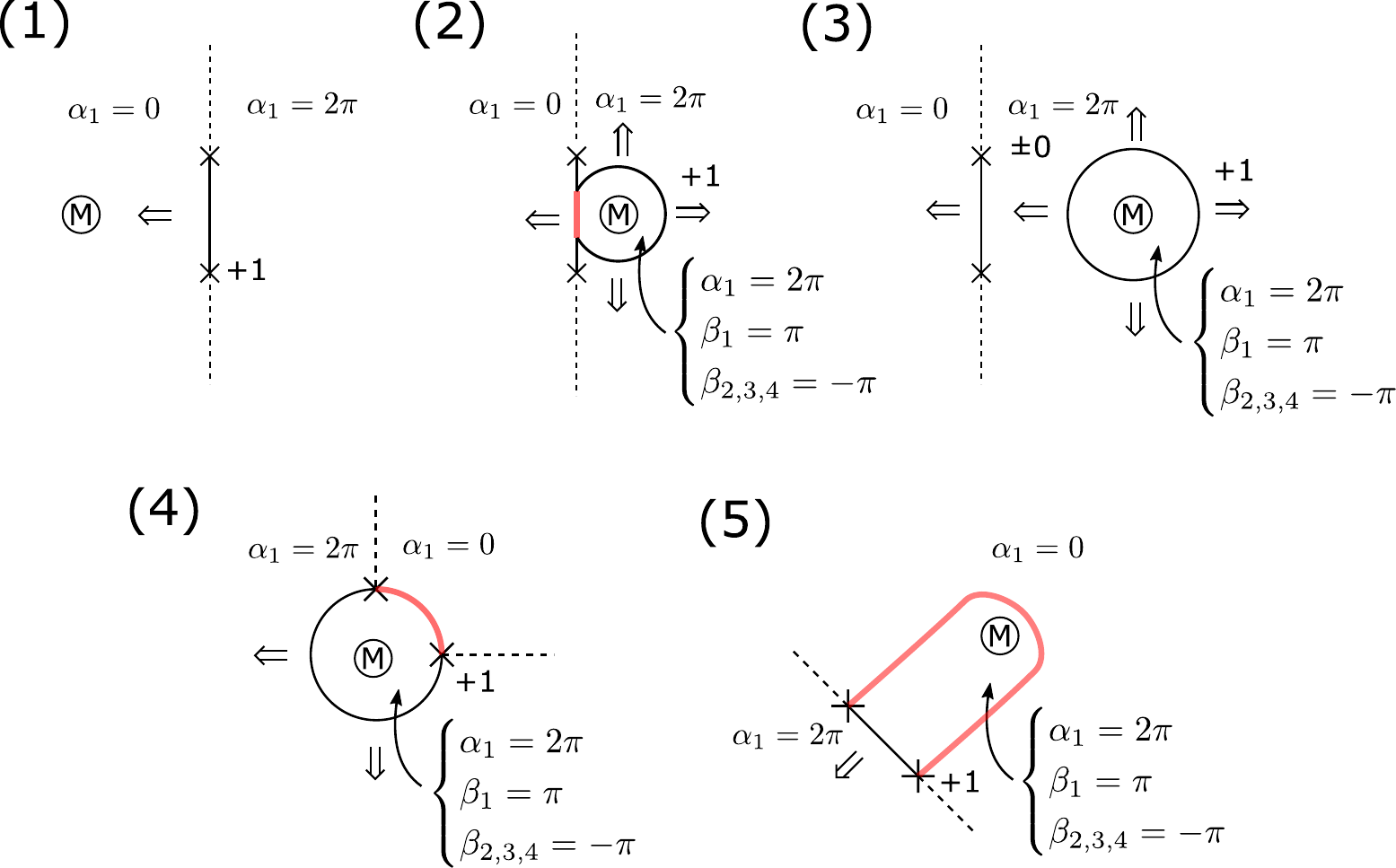}
\caption{
The soliton picture of the monopole scattering.
The black solid line shows the wall where $\alpha_1$ changes.}
\label{fig:pancake_monopole}
\end{figure}

We depict the evolution of the phases $\alpha_i$ and $\beta_i$ during the scattering in
\cref{fig:pancake_monopole}. We explain in the following the five
evolution steps in \cref{fig:pancake_monopole}.

\begin{enumerate}
  \renewcommand{\labelenumi}{(\arabic{enumi})}
  \item The wall where the phase $\alpha_1$ gradually changes from
  $\alpha_1=0$ to $\alpha_1=2\pi$ goes towards the monopole. The solid
  line represents the domain wall and its boundary is the string. The
  monopole is in the region where $\alpha_1=0$.

  \item When the wall reaches the location of the monopole%
  %
  %
  ,
  $\beta_i$'s change to satisfy the boundary conditions in
  Eq.~\eqref{boundary_cond} as
  \begin{align}
    \beta_1 = \frac{\alpha_1}{2},\quad 
    \beta_2 = \beta_3 = \beta_4 = -\frac{\alpha_1}{2}.
  \end{align}
  At the location of the monopole, $\alpha_1$ changes from $0$ to $2\pi$
  and thus $\beta_1$ changes from $0$ to $\pi$ while $\beta_{2,3,4}$
  change from $0$ to $-\pi$. Due to this change of $\beta_j$ near the
  monopole, there appears two types of walls that are connected: the
  wall where the value of $\beta_j$ changes with $\alpha_1$ fixed to be
  $2\pi$ (black line) and the wall where both $\alpha_1$ and $\beta_j$ change
  (red line).
  The latter one is in fact transparent because the values of the
  condensates do not change on the wall. Here the term ``transparent''
  means that there is no object actually. 
  %

  \item The string goes away from the
  monopole, and a monopole bag with charge 1 is left. Note
  that 
  the charge is completely transferred from the string to the bag. The neutral string is no longer stable and thus it disappears after the scattering.

  \item The wall can break by creating a string
  around which $\alpha_1$ changes from $0$ to $2\pi$. There also
  appears the transparent wall on which $\alpha_1$ and $\beta_{1,2,3,4}$
  change while satisfying the boundary condition.

  \item The wall representing the new particle remains as the final state.
  The quantum number is $(R,+1,\Box)$, which is the leftmost one in the
  bottom panel of \cref{fig:pancakes}.

\end{enumerate}

This description of the scattering is only schematic, and the actual process is not determined.
However, for consistency, the following two points must be satisfied.
(1) The helicity has to flip. This is consequence of the fact that the direction of the spin is determined by the direction of the wall.
(2) The bag configuration surrounding the monopole can deform continuously to the pancake. This deformation is the reverse of the process that the string shrinks and disappears, which is allowed when the charges are conserved under the process.

The new particle can go away from the monopole for an
arbitrary distance. 
Note, however, that there is no distance scale in
the theory when the fermions are massless. Therefore, there is no
actual meaning of far or close.

\subsection{The initial state with the opposite helicity}

We can also consider the case where the incoming particle is $\bar
b_j$, which is a right-handed particle with charge $+1$. In this case,
the total angular momentum is $\bm J = \hat{\bm n}$, where $\hat{\bm
n}$ is the direction of motion of the incoming particle, and therefore
the process cannot be described in the $s$-wave theory.
Because the initial state is not $s$-wave, the particle cannot interact with the monopole core and thus the helicity should not flip.
The soliton picture correctly reproduce this feature at least when the energy of the initial particle is large enough so that the string goes through the monopole.
In the process, the $U(1)$ charge moves from the bulk to the edge due to the Witten effect, and then the monopole bag with charge $-1$ and the pancake with charge $+2$ remain.
Because the edge state of the pancake has winding number $2$, the helicity of the pancake is $+1$, and thus the
total angular momentum is conserved. The flavor charge of the initial
state is $-[H_l]_{jj}$, and after the collision, the bag and the
pancake has flavor charges $[H_l]_{jj}$ and $-2[H_l]_{jj}$
respectively. The bag is deformed to the pancake corresponding to
$b_j$, which is a left-handed particle with charge $-1$. The pancake with charge $+2$ 
can collapse to two $\bar b_j$. 
The final state is $b_j+2\bar b_j$, which can be regarded as just a pair production.
A similar process in the skyrmion-monopole scattering is discussed in Ref.~\cite{Callan1984a}.

\section{Effects of fermion masses}

Let us introduce the mass terms
\begin{align}
  -m\varepsilon^{ab}(\chi_{1a}\chi_{2b} + \chi_{3a}\chi_{4b} + \mathrm{h.c.})
\end{align}
where $\chi_{ja}$ is a Weyl fermion with the flavor index $j$ and the $SU(2)$ index $a$.
For this mass term, the global symmetry reduces to $Sp(2)$ \cite{Strodthoff2012}.
In the effective theory, the mass terms are written as
\begin{align}
  -m(a_1b_2 - a_2b_1  + a_3b_4 - a_4b_3 + \mathrm{h.c.}),
  \label{mass}
\end{align}
where the minus sign in front of $a_2b_1$ and $a_4b_3$ is needed for the $SU(2)$ gauge invariance.
The $U(1)\times U(1)$ subgroup of $Sp(2)$ acts on the fermions as
\begin{align}
  &a_1\rightarrow e^{i\gamma_1}a_1,\quad a_2\rightarrow e^{-i\gamma_1}a_2\quad
  a_3\rightarrow e^{i\gamma_2}a_3,\quad a_4\rightarrow e^{-i\gamma_2}a_4,\notag\\
  &b_1\rightarrow e^{i\gamma_1}b_1,\quad b_2\rightarrow e^{-i\gamma_1}b_2\quad
  b_3\rightarrow e^{i\gamma_2}b_3,\quad b_4\rightarrow e^{-i\gamma_2}b_4.
\end{align}
Therefore, $\bar b_2$ has the same flavor charge as $a_1$.
This means that there is a candidate of the final state, and thus the puzzle disappears.
Therefore, we expect that the semiton states disappear when the mass term is introduced.
According to the numerical simulation \cite{Dawson1983}, the semiton state appears only as an intermediate state, and it decays into a usual fermion.
In the following, we give a four dimensional picture of this phenomenon.

In the massive case, there appears additional condensates:
\begin{align}
  \braket{a_1b_{2}} \simeq -\frac{\tilde c\mu}{r^2}, \quad
  \braket{a_2b_{1}} \simeq \frac{\tilde c\mu}{r^2}, \quad
  \braket{a_3b_{4}} \simeq - \frac{\tilde c\mu}{r^2}, \quad
  \braket{a_4b_{3}} \simeq \frac{\tilde c\mu}{r^2},
  \label{m_cond}
\end{align}
where $\mu$ is a renormalization scale and $\tilde c$ is a constant.
Let $\phi_1$, $\phi_2$, $\phi_3$ and $\phi_4$ be the phases of these operators.
Using $\alpha_j$ and $\beta_j$, they are expressed as
\begin{align}
    \phi_1 = \alpha_1 + \beta_2,\quad \phi_2 = \alpha_2 + \beta_1,\quad \phi_3 = \alpha_3 + \beta_4,\quad \phi_4 = \alpha_4 + \beta_3.
\end{align}
The s-wave analysis indicates that, in the effective theory of $\phi_j$, the mass term is written as
\begin{align}
  \frac{\mu}{4\pi r^2}\sum_{j=1}^4 (1-\cos\phi_j).
  \label{mass_term}
\end{align}

In the s-wave theory, a kink of $\phi_1$ describes a mixed state of $a_1$ and $\bar b_2$.
By analogy, in four dimensions, a pancake of $\phi_1$ should describe it.
Also in the massive case, the scattering of a monopole and a fermion can be described by using pancakes.

According to the s-wave analysis, the boundary condition at the core of the monopole is given as
\begin{align}
  \sum_j\partial_t\phi_j = \partial_t(\phi_1-\phi_2) = \partial_t(\phi_3-\phi_4)= \partial_r(\phi_1+\phi_2-\phi_3-\phi_4) = 0\quad \text{at }r=0,
  \label{m_boun}
\end{align}
where $r$ is the radial distance from the monopole.
This condition is obtained so that the $U(1)$ current $J$ and the flavor currents $J_l$ are conserved at the origin, which is the same as the massless case.
In contrast with the massless case, the values of the multi-fermion fields can change at the origin.
When the wall of $\phi_1$ traveling nearly at the speed of light reaches the core of the monopole, the values of $\phi_j$ gradually change toward
\begin{align}
  \phi_1 = \pi,\quad \phi_2=\pi,\quad \phi_3 = -\pi,\quad \phi_4 = -\pi,\qquad \text{at }r=0.
\end{align}
This gives the maximum of the mass term (\ref{mass_term}), and therefore the region where the values of $\phi_j$ given as this should decay into the vacuum.

The scattering of the s-wave fermion and the monopole can be calculated in the s-wave theory.
A classical numerical simulation was performed in the s-wave theory in Ref.\ \cite{Dawson1983}.
We can interpret the result as follows:
\begin{enumerate}
  \renewcommand{\labelenumi}{(\arabic{enumi})}
  \item Let the initial particle be $a_1$, which can be described as a kink soliton where $\phi_1(r=0)=0$ and $\phi_1(r=\infty)=2\pi$.
  \item In the region where $r\leq 1/m$, one can ignore the mass term, and thus the semiton state appears after the collision.
  In the bosonized theory, the state is described as a kink soliton where the values of the boson fields at the origin and the infinity are
  \begin{align}
    &\phi_1(r=0)=\phi_2(r=0)=\pi,\quad \phi_3(r=0)=\phi_4(r=0) =- \pi,\notag\\
    &\phi_1(r=\infty) = 2\pi,\quad\phi_2(r=\infty)=\phi_3(r=\infty)=\phi_4(r=\infty) =0.
  \end{align}
  \item Due to the mass term, the region where $\phi_1=\phi_2=\pi,\ \phi_3=\phi_4 =- \pi$ is no longer the vacuum, but it gives the maximum of the potential. Therefore, at some point of $r$ and $t$, the value of $\phi_j$ starts to change towards $\phi_j=0\bmod 2\pi$, which can be regarded as the pair creation of the semiton states $\bar b_1/2 + b_2/2 + b_3/2+b_4/2$ and $a_1/2 + \bar a_2/2 + a_3/2 + a_4/2$.
  Because the s-wave states of $a_j$ and $\bar a_j$ can only be the initial state, $a_1/2+\bar a_2/2 + a_3/2 + a_4/2$ goes back to the monopole.
  After the collision, it changes to $\bar b_2$ in the same way that $a_1$ changes to the semiton state, and thus the final state is $\bar b_2$.
  The pair of the semiton states is seen as a ripple of the boson field.
\end{enumerate}

\subsection{Interpretation in four dimensions}
\begin{figure}
\centering
\includegraphics[width=0.8\hsize]{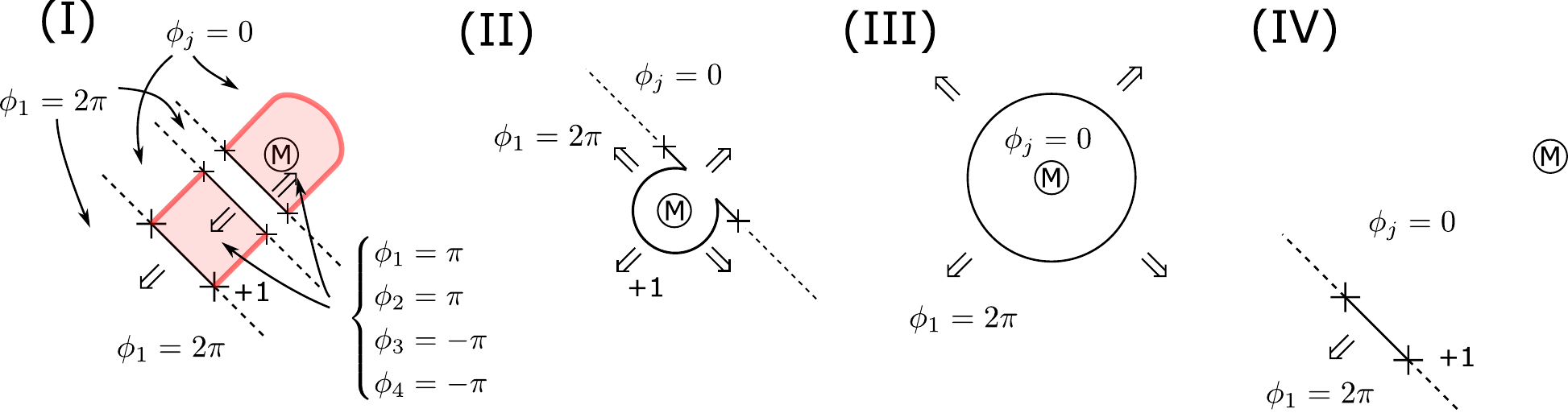}
\caption{The soliton picture of the monopole scattering for small fermion masses.
The red shaded region is not vacuum in the massive case.
}

\label{fig:pancake_monopole_massive}
\end{figure}

In the massive case, the field variables in the effective theory is $\phi_1$, $\phi_2$, $\phi_3$ and $\phi_4$ due to the additional condensates (\ref{m_cond}), which is different from the massless case.
For the spherically symmetric configuration, the same scattering process as the s-wave approximation is expected.
In the same way as the massless case, the fermions can be described as pancake solitons of $\phi_j$.
On the other hand, the semiton states cannot be described as isolated pancake solitons, but instead pancakes connected with the monopole by the region where $\phi_j=\pm \pi$.
As expected from the s-wave approximation, the pair creation of the semiton states occurs at some point.
See Fig.~\ref{fig:pancake_monopole_massive}.

\section{Summary}
\label{sec:summry}

The monopole catalyzed proton decay, called the Callan-Rubakov effect,
is a somewhat counter intuitive phenomenon. Even though the scale of
the baryon number violating interaction is much higher than that of
the proton mass or size, the rate of the proton decay around the
monopole is of order unity in the unit of the proton mass.
This effects have been used to put strong constraints on the monopole
abundance in the Universe by searching for neutrino flux from the
Sun~\cite{Arafune1983, Arafune1985, Kajita1985, Ueno2012} as well as
by the limits on the X-rays from neutron stars~\cite{Dimopoulos1982,
Kolb1984}.

Although the existence of the effect is theoretically robust with
massive fermions, there remains a puzzle in the interpretation of the
final states of the scattering processes in the limit of massless
fermions. Just by the conservation of quantum numbers, the final
states cannot be constructed from the superposition of the original
fermion states.
Logically, in this circumstance one may conclude that the massless
limit of QED with the monopole is not unitary, or something wrong in
the calculation.
In this paper, we find that there is no such pathology. The final
states of the scattering can actually be found as a form of solitons
made of the multi-fermion operators in the monopole background.

The solitonic object is a domain wall whose boundary is a loop of a
string. By the quantization of the edge mode on the string, fermionic
states can be found. The states indeed have the correct quantum
numbers for the missing final states. Moreover, the original fermions
can also be described as the solitons.
By using the soliton picture, one can describe the monopole-fermion
scattering as the evolution of the solitons. We can understand how the
chirality flipping happens and how the original particle deforms into
the new particle.
The massless QED seems to have a peculiar feature. The fermion
spectrum is doubled when there is a monopole somewhere!

By adding masses, $m$, to fermions, such new fermion states disappear.
In the soliton picture, the small mass deforms the particle into a
particle with a tail which stretches to the monopole. The object is
unstable when the distance between the particle and the monopole gets
further than $O(1/m)$.
The tail is detached from the monopole and eventually the
particle-tail system collapses into one of the original fermions. The
asymptotic states of the scattering are only the original fermions in
the massive case.
The ``new particle'' is an approximate picture of an intermediate
state that can only exist sufficiently near the monopole. In the
massless limit, everywhere gets sufficiently near the monopole.

Throwing an electron to the monopole turns the electron into a
pancake. Not only the soliton picture provides us with understanding
of the process, the discussion gives us a picture of how the
bosonization in two dimensions is lifted up to four dimensions.

\section*{Acknowledgements}
RK would like to thank Take Okui, Keisuke Harigaya, and John Terning
for collaborations and useful discussions in the early stage of this
work.
The work is supported by JSPS KAKENHI Grant No.~19H00689 (RK and RM),
and MEXT KAKENHI Grant No.~18H05542 (RK).

\bibliographystyle{JHEP_mendeley}
\bibliography{semiton}

\providecommand{\href}[2]{#2}\begingroup\raggedright\begin{thebibliography}{10}

\bibitem{Hooft1974a}
G.~{'t Hooft}, \emph{{Magnetic monopoles in unified gauge theories}},
  \href{https://doi.org/10.1016/0550-3213(74)90486-6}{\emph{Nucl. Phys. B}
  {\bfseries 79} (1974) 276}.

\bibitem{Polyakov1974}
A.M.~Polyakov, \emph{{Particle Spectrum in the Quantum Field Theory}},
  {\emph{JETP Lett.} {\bfseries 20} (1974) 194}.

\bibitem{Rubakov1982}
V.~Rubakov, \emph{{Adler-Bell-Jackiw anomaly and fermion-number breaking in the
  presence of a magnetic monopole}},
  \href{https://doi.org/10.1016/0550-3213(82)90034-7}{\emph{Nucl. Phys. B}
  {\bfseries 203} (1982) 311}.

\bibitem{Callan1982}
C.G.~Callan, \emph{{Dyon-fermion dynamics}},
  \href{https://doi.org/10.1103/PhysRevD.26.2058}{\emph{Phys. Rev. D}
  {\bfseries 26} (1982) 2058}.

\bibitem{Callan1983}
C.G.~Callan, \emph{{Monopole catalysis of baryon decay}},
  \href{https://doi.org/10.1016/0550-3213(83)90677-6}{\emph{Nucl. Phys. B}
  {\bfseries 212} (1983) 391}.

\bibitem{Kazama1977}
Y.~Kazama, C.N.~Yang and A.S.~Goldhaber, \emph{{Scattering of a Dirac Particle
  with Charge $Ze$ by a Fixed Magnetic Monopole}},
  \href{https://doi.org/10.1103/PhysRevD.15.2287}{\emph{Phys. Rev. D}
  {\bfseries 15} (1977) 2287}.

\bibitem{Jackiw1976}
R.~Jackiw and C.~Rebbi, \emph{{Solitons with fermion number 1/2}},
  \href{https://doi.org/10.1103/PhysRevD.13.3398}{\emph{Phys. Rev. D}
  {\bfseries 13} (1976) 3398}.

\bibitem{Dawson1983}
S.~Dawson and A.N.~Schellekens, \emph{{Monopole-fermion interactions: The
  soliton picture}},
  \href{https://doi.org/10.1103/PhysRevD.28.3125}{\emph{Phys. Rev. D}
  {\bfseries 28} (1983) 3125}.

\bibitem{Callan1984a}
C.G.~Callan and E.~Witten, \emph{{Monopole catalysis of Skyrmion decay}},
  \href{https://doi.org/10.1016/0550-3213(84)90088-9}{\emph{Nucl. Phys. B}
  {\bfseries 239} (1984) 161}.

\bibitem{Csaki2020}
C.~Csaki, S.~Hong, Y.~Shirman, O.~Telem, J.~Terning and M.~Waterbury,
  \emph{{Scattering Amplitudes for Monopoles: Pairwise Little Group and
  Pairwise Helicity}},  \href{https://arxiv.org/abs/2009.14213}{{\ttfamily
  arXiv:2009.14213}}.

\bibitem{Sen1983}
A.~Sen, \emph{{Conservation laws in the monopole-induced
  baryon-number-violating processes}},
  \href{https://doi.org/10.1103/PhysRevD.28.876}{\emph{Phys. Rev. D} {\bfseries
  28} (1983) 876}.

\bibitem{Callan1984}
C.G.~Callan, \emph{{The monopole catalysis S-matrix}},
  \href{https://doi.org/10.1063/1.34591}{\emph{AIP Conf. Proc.} {\bfseries 116}
  (1984) 45}.

\bibitem{Preskill1984}
J.~Preskill, \emph{{Magnetic Monopoles}},
  \href{https://doi.org/10.1146/annurev.ns.34.120184.002333}{\emph{Ann. Rev.
  Nucl. Part. Sci.} {\bfseries 34} (1984) 461}.

\bibitem{Rubakov1988}
V.A.~Rubakov, \emph{{Monopole catalysis of proton decay}},
  \href{https://doi.org/10.1088/0034-4885/51/2/002}{\emph{Rep. Prog. Phys.}
  {\bfseries 51} (1988) 189}.

\bibitem{Komargodski2018a}
Z.~Komargodski, \emph{{Baryons as Quantum Hall Droplets}},
  \href{https://arxiv.org/abs/1812.09253}{{\ttfamily arXiv:1812.09253}}.

\bibitem{Kitano2021}
R.~Kitano and R.~Matsudo, \emph{{Vector mesons on the wall}},
  \href{https://doi.org/10.1007/JHEP03(2021)023}{\emph{JHEP} {\bfseries 03}
  (2021) 23} [\href{https://arxiv.org/abs/2011.14637}{{\ttfamily
  arXiv:2011.14637}}].

\bibitem{Kogan1993}
I.I.~Kogan, \emph{{Kaluza-Klein and axion domain walls. Induced charge and mass
  transmutation}},
  \href{https://doi.org/10.1016/0370-2693(93)90877-K}{\emph{Phys. Lett. B}
  {\bfseries 299} (1993) 16}.

\bibitem{Maldacena1997}
J.M.~Maldacena and A.W.~Ludwig, \emph{{Majorana fermions, exact mapping between
  quantum impurity fixed points with four bulk fermion species, and solution of
  the “unitarity puzzle”}},
  \href{https://doi.org/10.1016/S0550-3213(97)00596-8}{\emph{Nucl. Phys. B}
  {\bfseries 506} (1997) 565}
  [\href{https://arxiv.org/abs/cond-mat/9502109}{{\ttfamily
  cond-mat/9502109}}].

\bibitem{Mandelstam1975}
S.~Mandelstam, \emph{{Soliton operators for the quantized sine-Gordon
  equation}}, \href{https://doi.org/10.1103/PhysRevD.11.3026}{\emph{Phys. Rev.
  D} {\bfseries 11} (1975) 3026}.

\bibitem{Coleman1975}
S.~Coleman, \emph{{Quantum sine-Gordon equation as the massive Thirring
  model}}, \href{https://doi.org/10.1103/PhysRevD.11.2088}{\emph{Phys. Rev. D}
  {\bfseries 11} (1975) 2088}.

\bibitem{Craigie1984}
N.~Craigie, W.~Nahm and V.~Rubakov, \emph{{Towards a complete QFT treatment of
  monopole induced baryon number violating transitions}},
  \href{https://doi.org/10.1016/0550-3213(84)90211-6}{\emph{Nucl. Phys. B}
  {\bfseries 241} (1984) 274}.

\bibitem{Kazama1983}
Y.~Kazama, \emph{{Condensates and the Boundary Condition in Monopole-Fermion
  Dynamics}}, \href{https://doi.org/10.1143/PTP.70.1166}{\emph{Prog. Theor.
  Phys.} {\bfseries 70} (1983) 1166}.

\bibitem{Callan1985}
C.~Callan and J.~Harvey, \emph{{Anomalies and fermion zero modes on strings and
  domain walls}},
  \href{https://doi.org/10.1016/0550-3213(85)90489-4}{\emph{Nucl. Phys. B}
  {\bfseries 250} (1985) 427}.

\bibitem{Witten1979c}
E.~Witten, \emph{{Dyons of charge e$\theta$/2$\pi$}},
  \href{https://doi.org/10.1016/0370-2693(79)90838-4}{\emph{Phys. Lett. B}
  {\bfseries 86} (1979) 283}.

\bibitem{Elitzur1989}
S.~Elitzur, G.~Moore, A.~Schwimmer and N.~Seiberg, \emph{{Remarks on the
  canonical quantization of the Chern-Simons-Witten theory}},
  \href{https://doi.org/10.1016/0550-3213(89)90436-7}{\emph{Nucl. Phys. B}
  {\bfseries 326} (1989) 108}.

\bibitem{Strodthoff2012}
N.~Strodthoff, B.-J.~Schaefer and L.~von Smekal, \emph{{Quark-meson-diquark
  model for two-color QCD}},
  \href{https://doi.org/10.1103/PhysRevD.85.074007}{\emph{Phys. Rev. D}
  {\bfseries 85} (2012) 074007}
  [\href{https://arxiv.org/abs/1112.5401}{{\ttfamily arXiv:1112.5401}}].

\bibitem{Arafune1983}
J.~Arafune and M.~Fukugita, \emph{{A limit on the solar monopole abundance}},
  \href{https://doi.org/10.1016/0370-2693(83)90810-9}{\emph{Phys. Lett. B}
  {\bfseries 133} (1983) 380}.

\bibitem{Arafune1985}
J.~Arafune, M.~Fukugita and S.~Yanagita, \emph{{Monopole abundance in the Solar
  System and the intrinsic heat in the Jovian planets}},
  \href{https://doi.org/10.1103/PhysRevD.32.2586}{\emph{Phys. Rev. D}
  {\bfseries 32} (1985) 2586}.

\bibitem{Kajita1985}
T.~Kajita, K.~Arisaka, M.~Koshiba, M.~Nakahata, Y.~Oyama, A.~Suzuki et~al.,
  \emph{{Search for Nucleon Decays Catalyzed by Magnetic Monopoles}},
  \href{https://doi.org/10.1143/JPSJ.54.4065}{\emph{J. Phys. Soc. Jpn.}
  {\bfseries 54} (1985) 4065}.

\bibitem{Ueno2012}
K.~Ueno, K.~Abe, Y.~Hayato, T.~Iida, K.~Iyogi, J.~Kameda et~al., \emph{{Search
  for GUT monopoles at Super–Kamiokande}},
  \href{https://doi.org/10.1016/j.astropartphys.2012.05.008}{\emph{Astropart.
  Phys.} {\bfseries 36} (2012) 131}
  [\href{https://arxiv.org/abs/1203.0940}{{\ttfamily arXiv:1203.0940}}].

\bibitem{Dimopoulos1982}
S.~Dimopoulos, J.~Preskill and F.~Wilczek, \emph{{Catalyzed nucleon decay in
  neutron stars}},
  \href{https://doi.org/10.1016/0370-2693(82)90679-7}{\emph{Phys. Lett. B}
  {\bfseries 119} (1982) 320}.

\bibitem{Kolb1984}
E.W.~Kolb and M.S.~Turner, \emph{{Limits from the soft X-ray background on the
  temperature of old neutron stars and on the flux of superheavy magnetic
  monopoles}}, \href{https://doi.org/10.1086/162645}{\emph{Astrophys. J.}
  {\bfseries 286} (1984) 702}.

\end{thebibliography}\endgroup
\end{document}